# Multi-property directed generative design of inorganic materials through Wyckoff-augmented transfer learning

**Shuya Yamazaki[1,3,#], Wei Nong[1,#], Ruiming Zhu[1,2,#],**

**Kostya S. Novoselov[3], Andrey Ustyuzhanin[3], Kedar Hippalgaonkar[1,2,3*]**

[1] School of Materials Science and Engineering, Nanyang Technological University, Singapore 639798, Singapore

[2] Institute of Materials Research and Engineering, Agency for Science, Technology and Research (A*STAR), Singapore 138634, Singapore

[3] Institute for Functional Intelligent Materials, National University of Singapore, Singapore 117544, Singapore

# These authors contributed equally

* Correspondence: kedar@ntu.edu.sg

# ABSTRACT

Accelerated materials discovery is an urgent demand to drive advancements in fields such as energy conversion, storage, and catalysis. Property-directed generative design has emerged as a transformative approach for rapidly discovering new functional inorganic materials with multiple desired properties within vast and complex search spaces. However, this approach faces two primary challenges: data scarcity for functional properties and the multi-objective optimization required to balance competing tasks. Here, we present a multi-property-directed generative framework designed to overcome these limitations and enhance site symmetry-compliant crystal generation beyond P1 (translational) symmetry. By incorporating Wyckoff-position-based data augmentation and transfer learning, our framework effectively handles sparse and small functional datasets, enabling the generation of new stable materials simultaneously conditioned on targeted space group, band gap, and formation energy. Using this approach, we identified $Cs_2Pt_3Se_7$, $Cd_2Ge_2O_3$, $Tl_3As_3S_4$, $Na_3MnSe_4$, $Al_6Ge_5S_{11}$, $Cd_3P_2Se_6$, $Rb_6Hg_2S_5$, and $Zr_2MnO_6$ as previously unknown thermodynamically and lattice-dynamically stable semiconductors in tetragonal, trigonal, and cubic systems, with bandgaps ranging from 0.13 to 2.20 eV, as validated by density functional theory (DFT) calculations. Additionally, we assessed their thermoelectric descriptors using DFT, indicating their potential suitability for thermoelectric applications. We believe our integrated framework represents a significant step forward in generative design of inorganic materials.

# Introduction

Materials discovery goes beyond finding new stable materials to designing novel materials with targeted functional properties for real-world applications. This property-directed materials design strategy is especially critical in fields like energy storage (batteries), and energy conversion such as thermoelectrics, and photovoltaic materials, where researchers continually push performance limits through enhanced non-equilibrium electronic, phononic and/or ionic transport. Here, additional data is required beyond ground state Density Functional Theory (DFT). Conventionally, property-directed material design relies heavily on either domain experience and knowledge, or more recently, via screening-based approaches, where computational methods such as DFT or machine learning surrogate models are employed to predict functional properties, filtering through large datasets to identify candidates that meet specific criteria. However, these methods are computationally intensive, especially when dealing with intractably vast search spaces.[1,2] Additionally, traditional materials screening methods are constrained by existing search spaces and researchers' empirical knowledge, limiting the range and efficiency of novel material discovery.

In the face of this material discovery dilemma, generative models have emerged as a powerful alternative, offering a more efficient route to materials discovery by directly including conditional sampling dependent on desired property. Different methods, such as variational autoencoders (VAEs), denoising diffusion probabilistic models (DDPMs), or flow-based models are employed to collapse materials into a lower-dimensional space or noise that follows a specific probabilistic distribution and revert them back to their original structures, allowing the models to learn the distribution of existing materials in relation to targeted properties.[3–7] Using various sampling strategies, generative models can explore the latent space to sample novel materials with preordained desirable properties, enabling a more focused and efficient search. For example, there have been some successes in such property-directed generative design in recent times. A semi-supervised variational autoencoder (SSVAE)[8], composed of three RNNs, effectively utilized both labeled and unlabeled data with variational inference to predict missing labels, enabling joint learning of property prediction and molecule generation from SMILES representations. It has successfully generated new molecular candidates with varied properties like molecular weight, hydrophobicity, and drug-likeness close to target values. Building upon this, Guided Diffusion for Inverse Molecular Design (GaUDI)[9] integrates an E(3)-equivariant graph neural network (GNN) with diffusion models to specifically target molecules with desired properties. It leverages the coarse-grained representation of molecules to efficiently explore chemical space by learning simpler distributions, enabling it to generate valid molecules that meet target criteria while improving performance and reducing computational demands in inverse design.

In inorganic crystal generative design, only a few initial studies have attempted to integrate property-guided design into their generative models. For instance, MatterGen[5] uses an SE(3)-equivariant graph-based diffusion model to generate materials by encoding crystal structures as graphs and employs adapter modules for property-guided generation through classifier-free guidance. This approach enables the model to generate novel materials with specific properties, like electronic bandgap and magnetic density, by fine-tuning property-conditioned scores with a limited labeled dataset. Conditional Crystal Diffusion Variational Autoencoder (Con-CDVAE)[10] combines diffusion models and variational autoencoders, using a two-step training method to first construct a latent space and then generate latent variables based on desired properties. It generates crystals with targeted attributes such as formation energy and band gap, though challenges remain in retaining space group symmetry and reducing discrepancies between initial generated structures and their DFT-relaxed structures. CrystalFormer[11], a space group-informed transformer model, adopts an approach where property prediction models are trained separately from a crystal generative model. It then combines a crystal probability prior with the property posterior, enabling plug-and-play conditional materials design through posterior Markov chain Monte Carlo sampling, while the demonstration of property-directed design is limited to one space group and lacks DFT validation of the generated structures or properties.

Despite the advancements made by these state-of-the-art generative models, controllable, property-driven generative design with customized constraints is not yet possible. In fact, no existing model has successfully demonstrated multi-property-directed design that conserves space group symmetry across all space groups. This is often limited by two primary challenges: data (size and quality) and multi-objective trade-off (for instance, property prediction and crystal generation). First, the functional property dataset is often small and sparse. This is primarily because computing target properties with DFT is computationally intensive and not always straightforward, while obtaining actual experimental values is even more laborious.[12] These constraints on data availability and quality directly impact the performance of generative models. Second, property-conditioned generation requires balancing between crystal generation and property prediction. When the target property becomes multi-objective, it introduces additional learning objectives to the model, which may degrade the prediction of other properties or the reconstruction of crystals. This trade-off makes it challenging for existing models to generate structurally valid symmetry-complaint crystals while simultaneously optimizing for specific properties.

Herein, we address these challenges directly via space group-informed data augmentation and transfer learning. Data augmentation refers to a set of techniques used to increase the size and diversity of training datasets, thereby enhancing model performance. In our case, given a dataset $D$ comprising training samples $S_1$ and corresponding labels $L_1$, data augmentation involves applying transformation operations $T_1$ to the original samples $S_1$ to generate new training data $S'_1$, while ensuring their corresponding labels $L_1$. In the case of inorganic crystal structure representation, training samples are crystal structures and labels are corresponding functional properties. These transformations, known as label-preserving operations, have been widely adopted in fields such as computer vision and are highly effective for expanding datasets, especially in domains where collecting large-scale data is expensive or challenging.[13] In materials science, where data acquisition is often expensive, a similar approach to data augmentation proves invaluable. Leveraging space group theory, label-preserving data augmentation can be effectively implemented through space group-constrained E(3) transformations. These transformations generate multiple symmetry-equivalent representations of the same crystal while preserving the structure-property relationship. Building on the previously featured Wyckoff representation[4], capturing the space group site symmetry of crystal structures, we developed Wyckoff-position-based data augmentation to overcome the limitations of small and sparce functional property dataset (Figure 1a). Furthermore, transfer learning has recently shown success in material property prediction by transferring knowledge from models trained on large datasets with one set of labels to smaller datasets with different labels, thereby improving model performance in data-limited domains.[14] We adopt a similar transfer learning approach, but in property-directed generative tasks, to enhance both the accuracy of property predictions and the validity of the generated materials.

In this work, we therefore introduce a versatile multi-property-directed symmetry compliant generative framework enabled by Wyckoff-based data augmentation and transfer learning. We demonstrate the impact on model performance across various properties in both forward predictions and conditional de novo generation tasks. Our model is further validated by the inverse design of novel and stable functional inorganic materials with a set of targeted properties and constraints. By balancing symmetry-preserving structure generation with property prediction, we advance toward a robust framework for AI-driven inverse design of functional inorganic materials with user-defined properties.

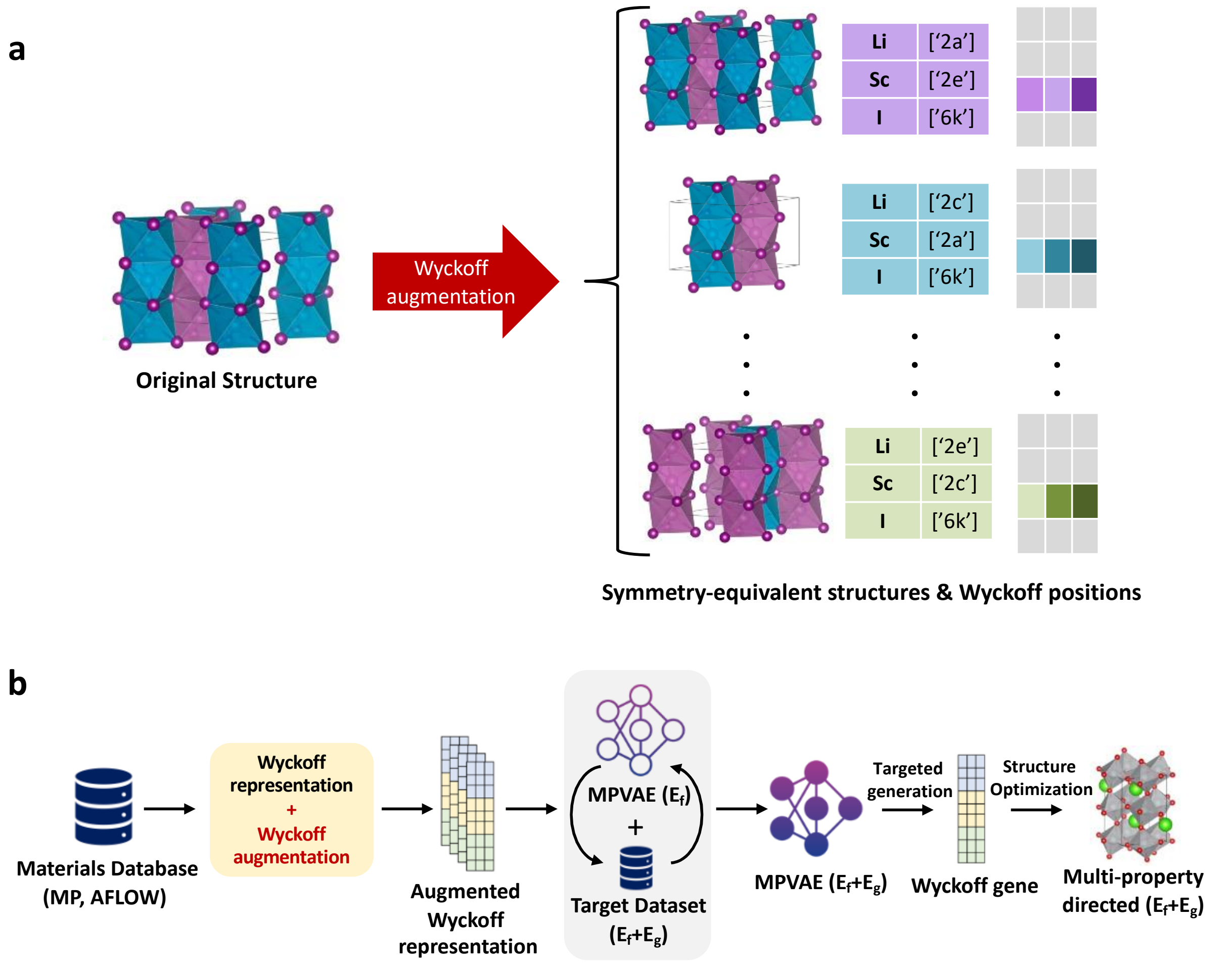

**Figure 1 | (a) Wyckoff augmentation and symmetry-equivalent structures.** Wyckoff augmentation applies space group-constrained E(3)-transformations to enrich crystallographic data by providing multiple, symmetry-equivalent views of the same structure. (b) An overview of the multi-property guided WyCryst generation framework. A multi-property directed WyCryst generation framework is proposed through Wyckoff augmentation and transfer learning. Crystallographic data undergoes Wyckoff augmentation, which is then converted into Wyckoff representations. The augmented representations, labeled with formation energy ($E_f$), are used to pretrain a Multi-Property directed Variational Autoencoder (MPVAE). The pretrained MPVAE is subsequently fine-tuned on a target dataset with multi-property labels (e.g., formation energy ($E_f$), band gap ($E_g$)). Finally, property-guided generation and structure optimization using DFT are performed.

# Results and discussion

The overview of the multi-property-guided inverse design framework based on the previous WyCryst[4] is described in Figure 1b. The original WyCryst generative framework was limited to single-property-directed design, but we extended this capability to accommodate multi-property-directed generation by implementing Wyckoff-position-based data augmentation and transfer learning (Figure 1b). The detailed architecture of the proposed generative framework is provided in the 'Methods' section.

In this section, we explore the potential of this integrated generation framework. We first test the performance of Wyckoff-augmented forward models on the AFLOW database[15] to demonstrate its effectiveness across different properties, comparing to several ML forward models. Then, we compare the performance of the original WyCryst to our proposed Multi-Property-directed Variational Autoencoders (MPVAE) enhanced by Wyckoff augmentation and transfer learning on the Materials Project (MP) database[16]. Finally, we perform property-directed de novo generation tasks using our integrated design framework. The trained MPVAE latent space is structured by desired properties, enabling the property-directed generation of novel materials. Additionally, we showcase and examine MPVAE-generated materials that not only possess desired properties but are also phonon stable, validated by the CrySPR workflows.[17]

## Forward Model Performance

We applied Wyckoff-position-based data augmentation to the original dataset and trained several structure-based forward models to predict multiple properties in the AFLOW dataset. The performance of these models is compared to state-of-the-art (SOTA) including CrabNet, a transformer-based self-attention model that learns inter-element interactions within compositions, and Roost, which models the similar relationships through a message-passing neural network for property prediction. [18–20] As part of our ablation studies on crystal structure-based generative models (WyCryst, FTCP[3]), we performed two tests: the model trained with original dataset as reference and the Wyckoff-augmented model, indicated by a "+" sign. FTCP, an invertible crystallographic representation, combines real-space (CIF-like) and reciprocal-space features and is coupled with a variational autoencoder with a property-learning branch. Table 1 shows the mean absolute error (MAE) performance metrics for benchmark material properties. The Wyckoff-augmented forward models, WyCryst+ and FTCP+, achieve the same order of magnitude accuracy compared against other SOTA models, therefore sufficient for generative design tasks. Notably, the Wyckoff augmentation reduces MAE of WyCryst forward models across all properties by 2-9%. Conversely, for FTCP, which incorporates both reciprocal and real space features, we applied Euclidean normalizers to the fractional coordinates as part of Wyckoff augmentation (see 'Wyckoff Augmentation' subsection in Methods). It is reported that applying random translations and rotations to fractional coordinates in the FTCP representation degrades the property prediction performance due to the lack of E(3)-invariance in the representation.[3] However, our results demonstrate that meaningful space group-constrained E(3)- transformations improve property predictions for certain functional properties, especially mechanical properties, highlighting the effectiveness of symmetry-informed data augmentation. This prompted us to test WyCryst-L+, an enhanced model that incorporates lattice parameters and fractional coordinates alongside the Wyckoff representation. WyCryst-L+ employs a dual augmentation strategy, combining Wyckoff letter-based and fractional-coordinate-based augmentations. This approach generates both symmetry-equivalent Wyckoff letters and their corresponding fractional coordinates, ensuring a more comprehensive representation of crystal structures. Interestingly, while adding structural features to Wyckoff representations alone did not enhance performance, this dual augmentation significantly improved forward model performance by 10-17% across all properties. This can be explained by the model's deeper understanding of space group symmetry rules in relation to properties, as further discussed in the following section.

This outcome aligns with expectations, as WyCryst-L+ utilizes minimal coarse-grained representations specifically designed for generative purposes, in contrast to more complex forward models fine-tuned exclusively for regression tasks on this dataset. Overall, the performance of WyCryst-L+ is comparable to that of dedicated forward models, and it provides sufficient accuracy for generative tasks. While we can further fine-tune WyCryst-L+ purely to improve the forward model performance, this might affect the reconstruction task in generative processes, which is our goal in this work.

**Table 1 | Benchmark results of forward models on the AFLOW dataset.** MAE scores of Roost, CrabNet, ElemNet, WyCryst, WyCryst+, WyCryst-L+, FTCP, FTCP+. "+" denotes the Wyckoff-augmented version of the representative models.

| Model | Bulk modulus (GPa) | Shear modulus (GPa) | Debye temperature (K) | Thermal conductivity ($Wm^{-1}K^{-1}$) | Thermal expansion ($\times 10^{-6}$ $K^{-1}$) |
|---|---|---|---|---|---|
| **Roost** | 8.82 | 9.98 | 37.2 | 2.70 | 3.96 |
| **CrabNet** | **8.69** | **9.08** | **33.5** | 2.32 | **3.85** |
| **ElemNet** | 12.1 | 13.3 | 45.7 | 3.32 | 5.42 |
| **FTCP** | 12.4 | 10.8 | 41.8 | **2.03** | 6.02 |
| **FTCP+** | 11.9 | 10.6 | 41.4 | 2.07 | 6.05 |
| **WyCryst** | 15.4 | 13.1 | 48.6 | 2.51 | 7.39 |
| **WyCryst+** | 14.7 | 12.2 | 47.8 | 2.38 | 6.71 |
| **WyCryst-L+** | 12.9 | 11.8 | 43.3 | 2.07 | 6.25 |

## MPVAE Generative Model Performance

Next, we evaluate the performance of MPVAE in generative tasks, with Wyckoff-augmentation and transfer learning. We processed the MP database into two datasets for the pre-training and fine-tuning of MPVAE:

(i) 40,426 MP ternary compounds with formation energy ($E_f$) labels (referred to as the 'source dataset'), and (ii) 16,480 MP ternary semiconducting compounds, a subset of the source dataset, having both formation energy ($E_f$) and non-zero band gap ($E_g$) labels (referred to as the 'target dataset').

Table 2 shows the property prediction and reconstruction performances of the original and modified WyCryst models. "WyCryst (single)" refers to the original single-property-directed PVAE model trained on the source dataset. In contrast, "WyCryst (multi)" represents the MPVAE model trained solely on the target dataset with two target properties simultaneously. This model faced a decline in overall performance, particularly in predicting $E_f$ and reconstructing the correct Wyckoff positions, which is expected as the target dataset with both properties ($E_f$ and $E_g$) has ~60% fewer datapoints. Such degradation is undesirable, impeding the generation of symmetry-compliant crystals: especially due to inaccuracies in Wyckoff position reconstruction linked to space group, which is an essential aspect of inverse design for inorganic crystals as symmetry defines property. We addressed these performance losses through 3 enhanced models: WyCryst+, TL-WyCryst, and TL-WyCryst+. The "+" signifies models that incorporate Wyckoff augmentation in the source and target datasets as described earlier, while "TL-" prefix indicates those trained using transfer learning with the MPVAE framework. WyCryst+ is trained exclusively on the Wyckoff-augmented target semiconducting materials dataset. TL-WyCryst undergoes pre-training on the source dataset before fine-tuning on the target dataset, and TL-WyCryst+ combines both approaches. Compared to the baseline WyCryst (multi) model, all three enhanced models show significant improvement in all property prediction and reconstruction tasks, as illustrated in Table 2. Firstly, since the source dataset is much larger than the target dataset, the pre-training process allows the model to learn fundamental space group symmetry rules from a broader range of crystal structures. In the subsequent transfer learning step, the model loads the pre-trained weights with the entire MPVAE encoder frozen, benefiting from the learned knowledge of symmetry-property relationships, rather than learning it from scratch on the limited target dataset. This two-step learning approach enables the pre-trained $E_f$ PVAE model to serve as a baseline model that can be fine-tuned with additional property labels on the smaller target dataset, which includes semiconducting materials with non-zero bandgap, $E_g$. Moreover, Wyckoff augmentation enhances the model's understanding of space group site-symmetry by allowing the model to learn the same crystal in multiple equivalent settings. This brings two key benefits: (1) by learning symmetry-equivalent Wyckoff representations, the model learns the higher-order symmetries and the fundamental building blocks of crystal structures better; (2) repeatedly learning the same crystal under different but equivalent settings encourages space group-constrained E(3)-invariance. These deeper understandings of underlying crystal symmetry, in turn, boost the prediction performance for both formation energy and band gap. As expected, the TL-WyCryst+ model achieved the largest improvements ($E_f$: +33.6%, $E_g$: +6.7%, Wyckoff: +32.5%) with better or comparable performance against single-property-directed models (Table 2). The enhancement in Wyckoff reconstruction accuracy is particularly important for inverse design, where generating correct Wyckoff positions for each space group is crucial to preserving the structure-property relationship.

**Table 2 | MPVAE generative model performance.** Property prediction (MAE) and reconstruction performance (accuracy) of enhanced WyCryst models compared with baseline WyCryst (single, multi) model. ”+” indicates the Wyckoff-augmented and "TL-" denotes the transfer-learned version of WyCryst.

| Model | $E_f$ (eV/atom) | $E_g$ (eV) | Element Accuracy (%) | Wyckoff Accuracy (%) | SG Accuracy (%) |
|---|---|---|---|---|---|
| **WyCryst (single)** | **0.063** | - | 99.6 | 91.3 | **90.8** |
| **WyCryst (multi)** | 0.113 | 0.448 | 97.1 | 64.1 | 86.1 |
| **WyCryst+** | 0.093 | 0.423 | 99.5 | 89.9 | 88.7 |
| **TL-WyCryst** | 0.077 | 0.442 | 99.7 | 87.6 | 90.6 |
| **TL-WyCryst+** | 0.075 | **0.418** | **99.9** | **96.6** | 90.1 |

## Multi-property-directed De Novo Generation Task

Using the best-performing TL-WyCryst+ generative model, we visualize the learnt latent space through Principal Component Analysis (PCA). By integrating the property-learning branch, the model learns the distribution of the training data with gradients of multiple target properties simultaneously, as shown in Figure 2(a-c). The property-learning branch of the TL-WyCryst+ model is trained on both formation energy and band gap, resulting in a "band gap + formation energy-structured" latent space. Interestingly, although this was not explicitly enforced, we also discovered that latent space shows a space group-related pattern in the form of crystal systems. With this structured latent space, one can perturb regions around the target properties of interest, allowing for property-conditioned sampling. This methodology can be applied to any properties or attributes, provided a sufficient dataset with appropriate labels and property-learning branches, showcasing the scalability and transferability of our approach.

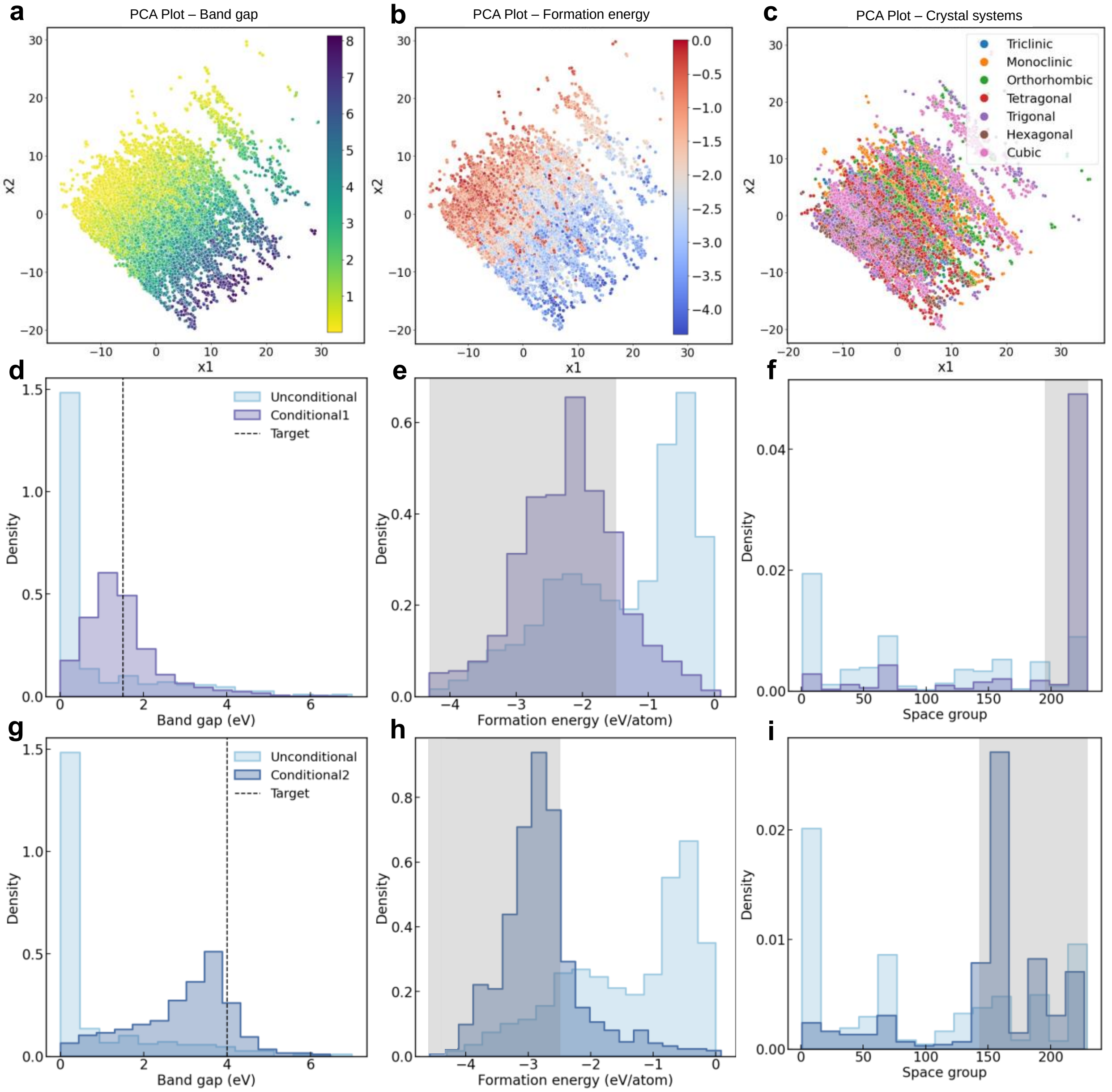

**Figure 2 | Multi-property-structured latent space and conditional generation.** Conditional sampling in the property-structured latent space enables controlling the distributions of multiple target attributes. (a-c) Property-structured latent space: (a) band gap (eV); (b) formation energy (eV/atom); (c) crystal systems; (d-i) Density of property values for conditional and unconditional generation using the MPVAE model for two different multi-property targets: Condition 1 (d) band gap = 1.5 eV, (e) formation energy < -1.5 eV/atom, and (f) space group ≥ 195; Condition 2 (g) band gap = 4.0 eV, (h) formation energy < -2.5 eV/atom, and (i) space group ≥ 143.

Here, we demonstrate the capabilities of the MPVAE to control the distributions of multiple targeted properties through conditional sampling in the property-structured latent space. Figure 2(d-i) shows the comparison of property-conditioned to unconditional sampling, where conditional sampling shifts the distribution of sampled compounds towards the predetermined target values or ranges. The light blue background distribution in Figure 2(d-i) represents property values from unconditional generation, reflecting the whole training distribution for each property. Using this distribution as a baseline, we tested two multi-property conditions: one with a band gap of 1.5 eV, formation energy < -1.5 eV/atom, and space group ≥ 195 (Figures 2(d-f)); the other with a band gap of 4.0 eV, formation energy < -2.5 eV/atom, and space group ≥ 143 (Figures 2(g-i)). These conditions are challenging, as they lie outside the training distribution and are applied simultaneously. Nevertheless, our MPVAE achieves out-of-distribution generation (light purple) compared to the training distribution (light blue). This indicates that by selecting seeds with a specific set of target properties, such as band gap, formation energy, and space group, and perturbing around those points, one can generate new crystals with the collection of multiple desired properties. Going beyond, generation can be further conditioned on specific chemical systems by selecting seeds comprising a targeted element set (e.g., Ca, Ti, O), as demonstrated in our earlier study. This approach enables the exploration of certain chemical systems or the restriction of generation to exclude precious metals or radioactive/toxic elements.[4] However, it should also be noted that imposing too many constraints may lead to broader deviations from the target values, as the sampling has to balance competing objectives while working with a smaller pool of reference materials.

Using the MPVAE with property-conditional sampling, we then performed multi-property-directed de novo generation. For this task, we set the aim as stable inorganic crystals with multiple targets and constraints: (i) semiconducting materials with band gap $0.5 < E_g < 2.0$, which is of interest for a wide range of applications such as photovoltaics, thermoelectrics, and optoelectronic devices; (ii) formation energy constraint with $E_f < -0.5$ eV/atom, as a proxy for stability; (iii) symmetry constraint (trigonal, hexagonal, cubic) with space group number ≥ 143. This design case is technologically relevant, making it an ideal testbed for our approach – essentially, we're asking our model to find thermodynamically favorable compounds that have a target bandgap with higher symmetries, and therefore a higher chance for being experimentally realized. We sampled 1,299 target-satisfying existing compounds from the training distribution as starting seeds in the property-structured latent space. From each reference seed, we perturbed and generated 20 reconstructed Wyckoff representations (hereinafter referred to as 'Wyckoff genes') per seed in the multi-property-structured latent space. After discarding existing and invalid ones, 8,082 novel and valid Wyckoff genes were yielded. All Wyckoff genes went through a series of filtration steps to ensure their physicochemical validity, which is illustrated in Figure 3a. Initially, the generated 8,082 novel Wyckoff genes are filtered based on the following sequential criteria: (1) 'true' label for SMACT charge neutrality[21], as evaluated from the accompanying compositions; (2) synthesizability score (SC)[22] > 0.4, and metallic score (MC) < 0.5, a value predicted by a binary classifier that predicts metals with scores close to '1' and non-metals near '0' based on the MP database; (3) exclusion of compounds containing bi-chalcogen, bi-halide, and chalcohalide (any two from O, S, Se, Te, F, Cl, Br, and I), radioactive elements and f-block metal elements; (4) a maximum of 50 atomic sites in the standard conventional cell; and (5) CHGNet $E_{\mathrm{hull}}$ $\leq 0.1$ eV/atom for the relaxed structures, which is referenced to structures from the MP database at the same level of energy prediction (Supplementary Figure 1, details in Methods section). We ended up with 135 candidates for further DFT calculations to check for property validation.

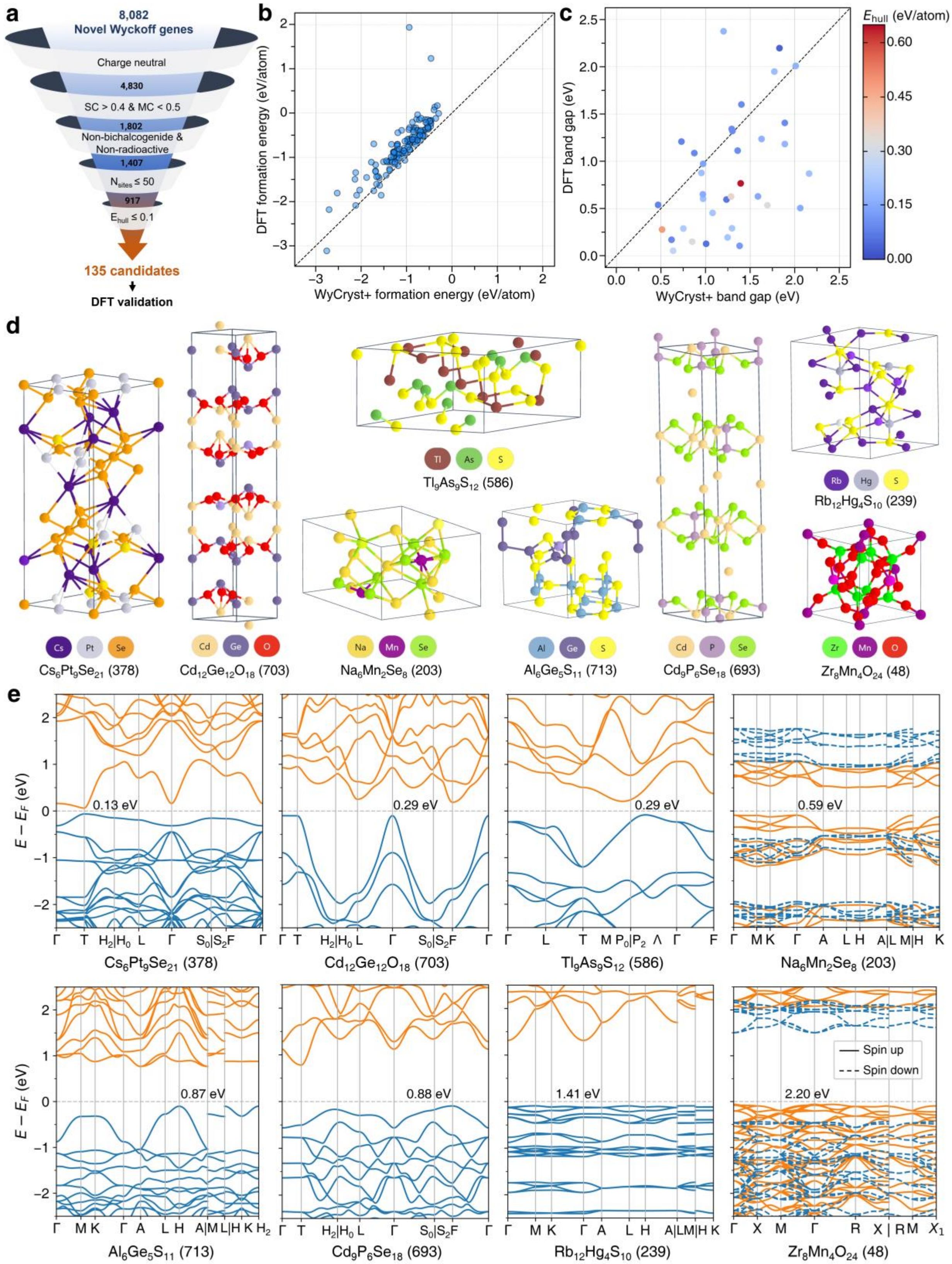

**Figure 3| Validation of property-directed design of semiconductor materials.** (a) Screening steps for generated compounds. (b) WyCryst+ predicted and DFT-calculated formation energies for 135 candidates. (c) WyCryst+ predicted and DFT-calculated band gaps for 35 semiconductors, colored by the DFT-calculated $E_{hull}$. (d, e) DFT-relaxed crystal structures and corresponding electronic band structures of eight generated semiconductors with Γ-phonon stability. The crystal structures are presented in the conventional cell, while the band structures are in the corresponding primitive cell. The numbers in the parentheses are the identifiers in the current study. For spin-polarized bands structures, the spin-up and spin-down components are shown with solid and dashed lines, respectively. The Fermi level is shifted to 0, which is obtained with a smearing width of 0.01 eV at 0 K.

We conducted detailed validation and investigation for 135 screened compounds using DFT relaxations, as well as phonon and electronic band structure calculations. As indicated in Figure 3b, WyCryst+ is capable of predicting the formation energies for unseen compounds solely from Wyckoff genes with a reasonable precision, showing an overall shift from the diagonal equity line compared to DFT calculated results. Among 135 candidates, however, 74% are found to have zero band gaps when calculated by DFT, despite being predicted to have finite band gaps by MPVAE. This issue is not unique to our model but is prevalent across other state-of-the-art surrogate models, with MEGNet mispredicting an energy gap in 76% of non-semiconductor materials as having an energy gap, as shown in Supplementary Figure 2. [23,24] We primarily attribute this to one fundamental issue; some compounds are labeled as having a zero band gap due to their Fermi level crossing either the conduction band or the valence band, despite having an energy gap either in the lower or higher energy levels within the band structure. Since the model does not account for the position of Fermi level, it learns only the energy gap value and we misclassify this as a zero band gap since the Fermi level is not within the bandgap. We attempted to address this issue using a Metal Classification (MC) model, which achieved 93.3% precision and 90.0% recall at a threshold of 0.5. However, the model could not differentiate between (1) the absence of an energy gap and (2) an energy gap in the deeper antibonding and bonding orbitals (or density of states), as both scenarios were labeled as '1' (metals). Despite these challenges, finding 35 previously undiscovered stable semiconductors that are not in the training set already counts as a success for the generative model. The band gaps and $E_{hull}$ for these 26% (35) valid semiconductors are listed in Supplementary Table 1. The corresponding WyCryst+-predicted and DFT-calculated band gaps are compared as plotted in Figure 3c, implying that WyCryst+ over-predict band gaps. This might be due to the absence of lattice parameters in the Wyckoff representation, due to which the interactions could be overly localized, so that the overlap between atomic orbitals is too weak to form continuous bands, resulting in the band gap overestimation.

For these 35 semiconductors, phonon modes and $E_{hull}$ results were analyzed to assess their dynamic and thermodynamic stability. In this study, a crystal is Γ-phonon stable if there is no imaginary mode at the Γ q-point, and is thermodynamically stable if the $E_{hull}$ is no more than an empirical threshold of 0.1 eV/atom. This leads to eight semiconductors with Γ-phonon stability and space group number ≥ 143, as highlighted in Supplementary Table 1. For these Γ-phonon stable crystals, the effective electron and hole masses were carefully analyzed from their band structures using the simplified form of Kane's dispersion relation,[25] as listed in Table 3. Additional critical features of band edges are also given. Detailed discussion of these features is provided in a later section.

The structures of the eight designed crystals are illustrated in Figure 3d. Intriguingly, there are structure analogs found from the Materials Project (MP) in each corresponding chemical system, as listed in Supplementary Table 2. The MP analogs bear similarity to the 8 WyCryst+-designed semiconductors especially with regards to the coordination environments or bond connections. Most of the WyCryst+-designed materials bear the same space group symmetry to the corresponding MP analogs. Significantly, all the MP analogs are either exactly on, or near, the convex hull and have band gap of > 1 eV, and most have been synthesized experimentally, accompanied with IDs from the Inorganic Crystal Structure Database (ICSD).[26] The low $E_{\text{hull}}$ in Table 3 implies their thermodynamic stability and experimental feasibility. The comparison for crystal structures and Wyckoff positions in Supplementary Figure 3 indicates that in some cases it appears that WyCryst+ performs a materials generation process involving (but not limited to) the following structural operations: substitution/replacement, swapping, insertion and deletion of Wyckoff positions onto the existing MP analogs. Such operations could be referred to as “local operation” within a specified space group symmetry. Intriguingly, WyCryst+ also appears to perform other operations beyond such traditional strategies, which can be referred to as “global operation” across different space group symmetries. For instance, there are no MP analogs for $Al_6Ge_5S_{11}$, and no Al-Ge-S ternaries in the MP database. Also, Supplementary Figure 4 shows that the WyCryst+-designed structure has similar local coordination environments but a different space group from the MP analog. Despite similarity to the existing structures, the design strategy of WyCryst+ is therefore not a simple elemental substitution but a collection of Wyckoff site occupancies allowable for a particular space group, while accounting for chemical complexity.

This further highlights a feature of the generative sampling strategy: it leverages information from neighboring stable, symmetry-compliant crystals in the latent space, thereby ensuring the generation of symmetry-obeying structures rather than those limited to P1 (translational) symmetry. However, this reliance on local perturbation-based sampling may favor known structural types and, as a result, may not be optimal for discovering entirely new structure types for targeted properties. To address this, one potential solution could be to explore property gradients within the latent space, using these gradients to guide sampling for both property optimization and structural novelty. This approach would allow for targeted exploration of new structure types within the property-structured latent space, independent of existing structures as starting points. However, due to the high dimensionality and complexity of our current latent space, this property-driven latent-space optimization poses a more advanced multi-objective challenge, which we reserve for future work.

## Thermoelectric Descriptors of the Inverse-designed Crystals

Next, the band structure-based descriptors that are useful for thermoelectric applications are analyzed. It is well-known that the PBE functional typically underestimates the band gap while generally it is good enough to describe the dispersion of bands, and it is also the reference functional on which the training data from the Materials Project are based, therefore, the band gaps and band structures from PBE functional are adopted for subsequent discussions. As listed in Table 3, all eight semiconductors are chalcogenides with DFT band gaps from 0.13 to 2.20 eV and favorable thermodynamic stability near the convex hull with DFT $E_{hull}$ ranging between 0.03 and 0.23 eV/atom. Among them, only $Cs_6Pt_9Se_{21}$ (378) shows a direct band gap of 0.13 eV at the T point while the rest are all indirect. The two Mn-containing compounds are magnetic semiconductors, and for $Na_6Mn_2Se_8$ (203) the band gap of the spin-up channel (solid lines) is much lower than that of the spin-down channel (dashed lines), while for $Zr_8Mn_4O_{24}$ (48) band gaps of both channels are close while the overall valence band maximum (VBM) and the conduction band minimum (CBM) occur for different spin channels (Figure 3d). Notably, $Al_6Ge_5S_{11}$ (713), $Cd_9P_6Se_{18}$ (693), and $Rb_{12}Hg_4S_{10}$ (239) with band gaps of 0.87, 0.88 and 1.41 eV could also potentially be explored for photovoltaic applications. As shown in Figure 3d, in the band structures of $Cs_6Pt_9Se_{21}$ (378), $Cd_{12}Ge_{12}O_{18}$ (703) and $Tl_9As_9S_{12}$ (586) and $Cd_9P_6Se_{18}$ (693) there exists a single isolated valence/conduction band near the Fermi level with different dispersion features. The highest valence band of $Cs_6Pt_9Se_{21}$ (378) is relatively flat, while that of $Cd_{12}Ge_{12}O_{18}$ (703) is strongly dispersive. This difference also is supported by lighter carrier effective mass (Table 3) in the band edges of $Cd_{12}Ge_{12}O_{18}$ (703), giving rise to the lightest hole effective mass (−0.19 $m_0$, $m_0$ is the free electron mass).

**Table 3 | Evaluation of DFT band structures**

| ID[a] | Space group | Formula | $E_{hull}$ (eV/atom) | $E_g^{DFT}$ (eV) | $m_h^*$ ($m_0$)[b] | $N_{VBM}$[c] | $m_e^*$ ($m_0$) | $N_{CBM}$ |
|---|---|---|---|---|---|---|---|---|
| 378 | 166 (R-3m) | $Cs_6Pt_9Se_{21}$ | 0.03 | 0.13 | −0.63 (**T**→Γ) | 1×1 (2) | 0.90 (**T**→Γ) | 1×1 (2) |
| 703 | 148 (R-3) | $Cd_{12}Ge_{12}O_{18}$ | 0.06 | 0.29 | −0.19 (**Γ**→L) | 1×1 (2) | 0.19 (**F**→$S_2$) | 3×1 (2) |
| 586 | 160 (R3m) | $Tl_9As_9S_{12}$ | 0.10 | 0.29 | −1.58 (**Λ**→Γ) | 2×2 (2) | 1.44 (**M**→T) | 2×1 (2) |
| 203 | 186 ($P6_3mc$) | $Na_6Mn_2Se_8$ | 0.06 | 0.59[d] | −1.63 (**M**→L) | 3×1 (1) | 0.60 (**A**→Γ) | 1×2 (1) |
| 713 | 143 (P3) | $Al_6Ge_5S_{11}$ | 0.22 | 0.87 | −0.73 (**H**→K) | 2×1 (2) | 0.85 (**A**→H) | 1×2 (1) |
| 693 | 148 (R-3) | $Cd_9P_6Se_{18}$ | 0.20 | 0.88 | −1.44 (**F**→$S_2$) | 1×1 (2) | 0.17 (**T**→$H_2$) | 1×1 (2) |
| 239 | 186 ($P6_3mc$) | $Rb_{12}Hg_4S_{10}$ | 0.23 | 1.41 | — | — | 0.47 (**Γ**→M) | 1×1 (2) |
| 48 | 205 (Pa-3) | $Zr_8Mn_4O_{24}$ | 0.20 | 2.20[d] | — | — | — | — |

[a] The index identifier for reference in the current study.

[b] The fitting directions near the band edges for the band effective mass are given in the parentheses and the k-points where the extrema occur are in bold; $m_0$ is the free electron mass.

[c] Total valley degeneracies are given in the form of $N_V = N_k \times N_b$, where $N_k$ is the degeneracy of k wavevectors in the first Brillouin zone where the iso-energy valleys locate at, as imposed by the crystal symmetry, and $N_b$ is the orbital (band) degeneracy. The spin

degeneracies are given in the parenthesis, (1) for spin-polarized and (2) for non-spin-polarized; — indicates the strong non-parabolicity of band edges so that the quasi-parabolic fitting fails, and there is no explicit valley.
[d] Values for the spin-up channel.

The first four identified semiconductors, i.e., $Cs_6Pt_9Se_{21}$ (378), $Cd_{12}Ge_{12}O_{18}$ (703), $Tl_9As_9S_{12}$ (586), and $Na_6Mn_2Se_8$ (203) might be promising for thermoelectric applications, considering three fundamentally coupled aspects: 1) low/medium band gaps required for facilitating a maximized Seebeck coefficient ($S$) at intermediate temperatures stemming from the Goldsmid-Sharp relation to facilitate optimized doping, and 2) light carrier (conductivity) effective masses ($m_c$*) that is required for the improved electrical conductivity ($\sigma$), and 3) high valley degeneracy ($N_V$) for the enhanced thermoelectric power factor ($S^2\sigma \sim N_V/m_c^*$).[27–30] Given the empirical "10 $k_BT$ rule" for band gaps[31], where $k_B$ and $T$ are the Boltzmann constant and absolute temperature, respectively, one can have an estimated band gap of < ~ 0.8 eV required for the thermoelectric material working below ~ 1,000 K. The $N_V$ of $Cs_6Pt_9Se_{21}$ (378) and $Cd_{12}Ge_{12}O_{18}$ (703) are relatively small due to the band extrema locating at the high-symmetry K-point, however, the small band effective masses ($m_b$*, Table 3) could also indicate relatively small $m_c$*, which facilitates a high mobility. Nevertheless, one should note that the anisotropy of $m_b$* along different directions also affects this indication. The two semiconductors can also be interesting when used as n-type thermoelectric materials as in such case the other valleys with close eigen energy also contributes to the carrier transport: for $Cs_6Pt_9Se_{21}$ (378) from the Γ electron valley with only 0.10 eV above the CBM and small $m_e$* (0.15 $m_0$), and for $Cd_{12}Ge_{12}O_{18}$ (703) from the L electron valley ($N_V$ = 3) with only 0.06 eV above the CBM and small $m_e$* (0.19 $m_0$). If the electron transport process at high temperature occurs within an energy window such that the valleys with near-equal energy are reachable, leading to band convergence via doping or alloying[30,32], $Cd_{12}Ge_{12}O_{18}$ (703) can give rise to a total $N_V$ of 6 near the CBM. Given that $Cd_{12}Ge_{12}O_{18}$ (703) gives rise to a DFT $E_{hull}$ of only 0.06 eV/atom, it might be suitable for experimental synthesis. For further consideration, however, a careful evaluation of the Fermi surface complexity factor is needed based on band gaps with proper precision.[43,45] As for $Tl_9As_9S_{12}$ (586), it has heavier carrier effective mass but larger $N_V$ (4 for Λ hole valley; 2 for M electron valley and 3 for F electron valley with 0.16 eV above the CBM). Similar cases happen to the magnet $Na_6Mn_2Se_8$ (203) with slightly heavy carries, where multiple hole valleys could contribute to the band convergence (Figure 3e).

# Conclusions

In conclusion, this work presents a multi-property-directed generative model for the inverse design of inorganic materials by integrating Wyckoff-position-based data augmentation and transfer learning. Our framework addresses key challenges in materials discovery, particularly data scarcity and multi-objective optimization, improving both symmetry-preserving crystal generation and prediction accuracy for multiple target properties like formation energy, band gap, and space group. The integration of Wyckoff augmentation and transfer learning enhanced both forward property predictions and crystal structure generation by leveraging space group site symmetry and the two-step learning approach. Furthermore, we showcased the MPVAE's capability of controlling the distribution of multiple target properties in our multi-objective de novo generation tasks. Notably, this framework successfully generated 8 novel semiconductor materials with targeted functional properties, thermodynamic stability, and lattice-dynamic stability, offering a significant step forward in AI-driven inverse design of inorganic materials.

# Methods

### Crystal symmetry, Wyckoff positions, and Euclidean normalizers

Inorganic crystalline materials can be characterized by their smallest repeating unit cell, which contains full symmetry of the crystal structure. Every structure possesses at least P1 symmetry stemming from its fundamental global lattice translation. Beyond this global translational symmetry, the internal symmetry within the unit cell can be further mathematically classified into 229 space groups based on their symmetry operations, such as rotations, reflections, translations, or combinations thereof within a unit cell. A subset of these space group operations forms the site symmetry group, which consists of symmetry operations that leave a specific point in the crystal invariant. Points in a crystal sharing the same site symmetry group are grouped into Wyckoff positions under a given space group. These positions are denoted by combinations of letters (e.g., a, b, c) and multiplicities (e.g., 2, 4, 8), offering a concise description of atomic positions within a unit cell with significantly fewer parameters while preserving essential symmetry information.

Based on group theory, multiple Wyckoff positions within a crystal can be symmetry-equivalent through higher-order symmetry operations beyond those defined by the site symmetry group.[33] This concept, known as the "symmetry of the symmetry," is formalized through the Euclidean normalizer of the space group. While site symmetry operations only maintain symmetry around a specific point, the Euclidean normalizer includes operations that map between different symmetry-equivalent Wyckoff positions, preserving the overall structure of the space group on a global scale. The Euclidean normalizer acts as a supergroup of the site symmetry group, allowing for mapping between distinct but equivalent Wyckoff positions and fully capturing the hierarchical structure of symmetries within a crystal.

### Wyckoff representation

Wyckoff representation consists of two key components: space group array $S_i$ and Wyckoff array $X_i$. The space group array is a one-hot encoded matrix with dimensions corresponding to the total number of space groups $\{230\}$. The Wyckoff array $X_i = (F_i, V_i, W_i)$ incorporates information on stoichiometry, atomic features, and Wyckoff sites occupancy. $F_i$ is a one-hot encoded stoichiometry matrix that symbolizes the crystal's chemical formula $\{A_l B_m C_n\}$. $V_i$ represents the atomic features matrix adapted from crystal graph convolutional neural network (CGCNN).[34] Lastly, $W_i$ indicates the Wyckoff site occupancy and multiplicity for each element. The stoichiometry matrix $F_i$ facilitates the reconstruction of the chemical formula, while the combination of $S_i$ and $W_i$ ensure the generated crystals conform to space group site symmetry rules.

### Wyckoff augmentation

We implement crystal data augmentation by utilizing Euclidean normalizers to enumerate all unique symmetry-equivalent crystal representations for a given space group. The augmentation process entails three elements: the space group, Wyckoff positions, and Wyckoff sets or coset representatives of Euclidian normalizers of a given space group (hereinafter referred to as normalizers). Wyckoff sets are defined as all points whose site-symmetry groups are conjugate subgroups of the normalizer $N$ of the space group $G$.[35] For instance, space group 2 has 8 normalizers that map Wyckoff positions onto symmetry-equivalent positions, forming 8 distinct Wyckoff sets. This effectively multiplies the data size 8-fold. Conversely, in space group 225, where only Wyckoff positions ('a', 'b') or ('h', 'i') are interchangeable, while other positions remain fixed when the normalizer is applied. In such cases, if no points in the crystal are located at any of these symmetry-equivalent positions, applying normalizers to these Wyckoff positions retains the original crystal representations. Hence, Wyckoff augmentation is performed only if the crystal has Wyckoff positions with equivalent positions under its given space group. As a result, the number of equivalent representations is equal to or fewer than the number of normalizers.

To implement Wyckoff augmentation, we extracted all Wyckoff sets and corresponding normalizers for each space group from the Bilbao Crystallographic Server.[36] We then constructed dictionaries of all unique Wyckoff sets and normalizers. The Wyckoff sets dictionary includes sets of symmetry-equivalent letters for each space group, while the normalizer dictionary contains 3×4 matrices. In these matrices, the 3×3 component applies rotations, reflections, and inversions, while the 3×1 component applies translations to the fractional coordinates. Next, we developed an algorithm to systematically enumerate all possible symmetry-equivalent Wyckoff positions or corresponding fractional coordinates for each crystal within its given space group. This algorithm generates augmented crystal representations based on either Wyckoff sets or fractional coordinates. We employed two distinct approaches for augmentation: (1) using the Wyckoff sets dictionary to augment Wyckoff-based representations, and (2) applying a normalizer from the normalizer dictionary to the fractional coordinates to produce symmetry-equivalent coordinates. The latter approach is particularly useful for structure-aware models that do not inherently incorporate Wyckoff positions in their representation and instead rely on fractional coordinates for direct positional encoding.
Leveraging Wyckoff augmentation, we then enrich crystallographic data by providing multiple, symmetry-equivalent views of the same structure, as illustrated in Figure 1a. These equivalent representations, which retain the underlying structure and properties, enable label-preserving data augmentation in crystallographic datasets. This augmented information can potentially enhance the performance of machine learning models in predicting material properties and generating crystal structures by enabling the model to learn the higher-order symmetries of space group operations. However, it is also critical to consider the compatibility between this augmentation method, the representations used, and the model itself, as it affects the overall effectiveness of the approach.

**MPVAE Model**

For crystal generation tasks, we used Variational Autoencoders (VAEs) with the latent space organized by specific target properties. Building on the WyCryst Property-directed Variational Autoencoders (PVAE)[4], where the encoder and decoder are based on Convolutional Neural Networks (CNNs), we introduce Multi-property-directed Variational Autoencoder (MPVAE). In MPVAE, property-learning branches connect the latent space to several target properties via fully connected layers. This architecture enables the model to learn the distribution of crystal structures in relation to multiple physical or chemical properties, such as formation energy and band gap. The encoder parameterizes a multivariate Gaussian distribution in the latent space by outputting $Z_{mean}$ and $Z_{variance}$, allowing for smooth sampling and generation of new materials.
The MPVAE incorporates four loss functions: (1) Reconstruction loss $L_{recon}$, a combination of mean squared error (MSE) for the Wyckoff array $X_i$ and cross-entropy loss for the space group array $S_i$, ensures the model accurately reconstructs the input Wyckoff representations; (2) KL divergence loss $L_{KL}$, shapes the latent space into a multivariate Gaussian distribution to regularize it and ensure continuity; (3) Property loss $L_{prop}$, defined as the MSE between the true and predicted values of multiple target properties, guides the property-learning branches to capture the relationship between crystal structures and physical properties; (4) Wyckoff loss $L_{Wyckoff}$, ensures that space group site symmetries are preserved by minimizing the MSE between the original and reconstructed formulas, with a focus on Wyckoff sites.

Balancing these losses, especially for multi-property learning, poses challenges in maintaining both accurate property prediction and symmetry-preserving crystal reconstruction. To address this, we implemented Wyckoff augmentation and transfer learning techniques, enhancing the model's ability to generate symmetry-compliant materials while steering them toward multiple targeted properties.

We trained our MPVAE in two steps using transfer learning due to the limited size of the labeled dataset with target properties. First, we pre-trained our model on the entire source dataset. Only formation energy ($E_f$) was included in the property loss function during pre-training. Next, we fine-tuned the pre-trained model using the target dataset, which is a subset of the source dataset with both $E_f$ and band gap energy ($E_g$) labels. During this phase, we froze the entire MPVAE's encoder, and all batch normalization layers in both the encoder and decoder to keep them in inference mode. In the fine-tuning process, we included both formation energy and band gap in the property loss functions.

**Conditional and Unconditional Sampling**

For conditional sampling, we sample from the training distribution within a target property range, allowing for the model's property prediction error tolerance. In contrast, for unconditional sampling, we randomly sample from the entire training distribution. In both approaches, we set these seeds as reference points in the latent space and apply a local perturbation method using Gaussian noise to sample around them. This sampling strategy, combined with the property-structured latent space, enables us to discover novel materials within specific target property ranges.

**Validation Procedures**

Initial structures for Wyckoff genes are generated using PyXtal.[37] Those initial structures are firstly relaxed by a universal machine learning interatomic potential, CHGNet[38], via the interface in CrySPR[17]. CHGNet predicted total energies of these relaxed structures were then used to get the energy above the hull ($E_{\text{hull}}$). The reference convex hulls are based on on-the-hull crystal structures from the corresponding chemical systems in the MP database, with their total energies also calculated using CHGNet. The CHGNet $E_{\text{hull}}$ is obtained using the PhaseDiagram module in pymatgen.[39] The CHGNet-relaxed structures are used as the input to a DFT workflow, as described in our previous work.[4]

In the automated DFT workflow, structure relaxation, electronic band structures, density functional perturbation theory (DFPT) were performed using the Vienna ab-initio simulation package (VASP) with the plane-wave basis set.[40] The electron-ion interaction is described by the projector augmented wave (PAW) pseudo-potentials.[41] The exchange-correlation of valence electrons is described using the Perdew-Burke-Ernzerhof (PBE) functional within the generalized gradient approximation (GGA).[42] The kinetic energy cutoff was set to 520 eV. Convergence tolerances of $10^{-8}$ eV for total energy and $10^{-4}$ eV Å$^{-1}$ atom$^{-1}$ for force were used. The Monkhorst-Pack scheme[43] is used to sample k-points in the Brillouin-zone. Γ-centered k-meshes with spacing of 0.15 Å$^{-1}$ was used for both structure relaxation and DFPT calculations, while spacings of 0.10 Å$^{-1}$ were employed for static runs. The intersection between two high-symmetry k-points was set to 40 for the band structure calculations. The tetrahedron method with Blöchl corrections[44] is employed for orbital occupancy for self-consistent field (SCF) calculations for obtained ground-state total energies and charge densities, while the Gaussian smearing with width of 0.01 eV is used for calculations for structure relaxation and band structures. The simplified DFT+$U$ approach proposed by Dudarev et al.[45] was employed in the calculations only for the oxides and fluorides that contains one or more of the following transition metals: Co ($U$ = 3.32 eV), Cr ($U$ = 3.7 eV), Fe ($U$ = 5.3 eV), Mn ($U$ = 3.9 eV), Mo ($U$ = 4.38 eV), Ni ($U$ = 6.2 eV), V ($U$ = 3.25 eV), W ($U$ = 6.2 eV), consistent with the Materials Project.[46] Spin-polarized relaxations initialized with ferromagnetic, high-spin valence configurations were also performed to check if there is any magnetic atom with magnetism ≥ 0.15 $\mu_B$. The band structures were calculated along the high-symmetry k-path as generated using SeeK-path.[47] The band structures and band gaps in the current study are reported by the aforementioned DFT settings.

To use the MP convex hull as the reference hull, additional DFT relaxations and SCF calculations using the VASP settings from MPRelaxSet and MPStaticSet in pymatgen were further performed based on the previously relaxed structures. The raw total energies are then corrected using the correction scheme of MaterialsProject2020Compatibility in pymatgen before putting into the PhaseDiagram to obtain the DFT formation energies and DFT $E_{\mathrm{hull}}$ with comparable settings. Should be carefully noted that the precision parameters, generated by MPRelaxSet, are too coarse compared with those set in previous relaxations, especially thresholds for convergence (~ $2 \times 10^{-4}$ eV for energy, and ~ $2 \times 10^{-3}$ eV for ionic relaxations while without setting for Hellmann-Feynman forces for each atoms) and the density of k-meshes (equivalent to a spacing of only ~ 0.35 $Å^{-1}$). They are not strictly appropriate for structure relaxations for new generated structures that typically might be far off equilibrium with large Hellmann-Feynman forces on atom and stresses on cell, therefore fine settings are needed for minimization to find the ground-state equilibrium configurations.

The band effective masses of carriers for the single valley at the band extrema, $m^*$, are evaluated based on the simplified form of Kane's (quasi-parabolic) dispersion relation[48,49] near the corresponding band edges, as given by $\hbar^2 k^2/(2m^*) = E(1 + \alpha E)$, where $\hbar$, $k$, $E$, and $\alpha$ are the reduced Planck constant, wavevector, eigen energy, and non-parabolicity parameter, respectively. The extraction of the inertial effective masses was implemented using the *effmass* package.[50]

**Data and Computation**

The datasets are obtained from the MP[16] and AFLOW[15] database respectively. We queried a total of 66,643 ternary compounds from the MP database accessed on 4 July 2023 and 4,905 compounds from the AFLOW database. We run all the model training and data sampling on a server with the following configuration: 2x Intel Xeon Gold 6336Y (24 cores/each) CPU, 256GB DDR4 (32GB×8) RAM, 2 TB SSD, and NVIDIA A40 GPU with 10752 CUDA cores and 48GB VRAM. We run DFT calculations on a server platform with the following configuration: Dual socket AMD EPYC 7713 64-cores @2.0 GHz, 512 GB DDR4 RAM.

## DATA AVAILABILITY

The datasets used for training, testing, and validation were obtained from the Materials Project database accessed on 2023.7.4. The source code for the MPVAE model, Wyckoff augmentation, and transfer learning can be found at https://github.com/shuyayamazaki/WyCryst-P. For Wyckoff Gene post-processing using CrySPR, please refer to https://github.com/Tosykie/CrySPR.

## ACKNOWLEDGEMENTS

K.H. acknowledges funding from the MAT-GDT Program at A*STAR via the AME Programmatic Fund by the Agency for Science, Technology and Research under Grant No. M24N4b0034. K.H. also acknowledges funding from the NRF Fellowship NRF-NRFF13-2021-0011. The computational work of DFT calculations for this article was fully performed on resources of the National Supercomputing Centre, Singapore (https://www.nscc.sg) under the project No. 12003663.

## AUTHOR CONTRIBUTIONS

K.H. conceived the research. S.Y. conceptualized, developed, and refined the Wyckoff augmentation, transfer learning, and MPVAE model, under the guidance of R.Z. and W.N. S.Y. performed screening and CrySPR post-processing of generated Wyckoff genes. W.N. conducted and analyzed the DFT calculations for crystal structure refinement and property calculations. K.H., S.Y., W.N., and R.Z. wrote the manuscript, with input from all co-authors.

## COMPETING INTERESTS

K.H. owns equity in a startup focused on using machine learning for materials discovery.

# Supplemental Information

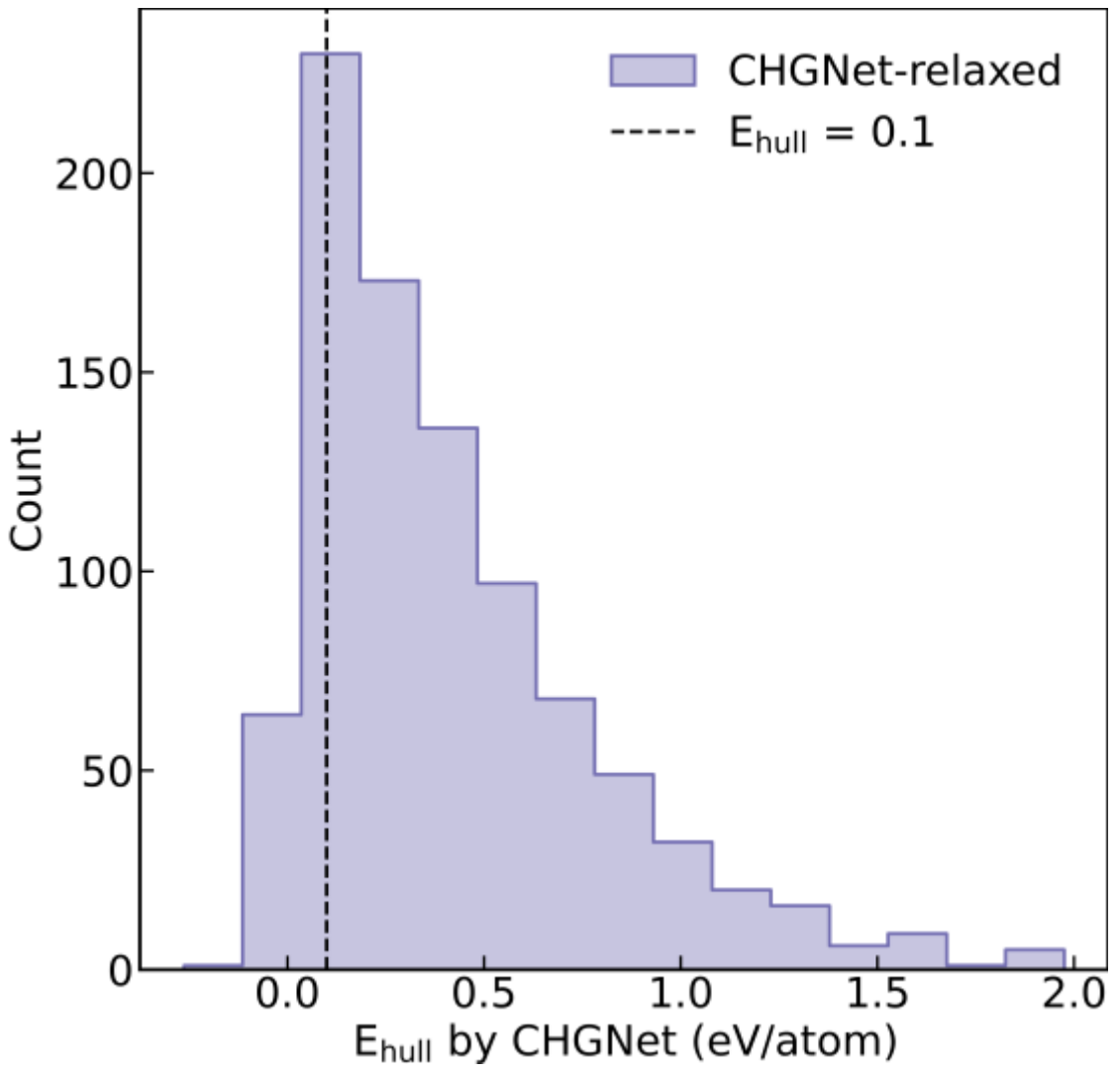

**Supplementary Figure 1 | Distribution of energy above hull ($E_{hull}$) of 917 generated crystals.** The results are given at the CHGNet calculation level.

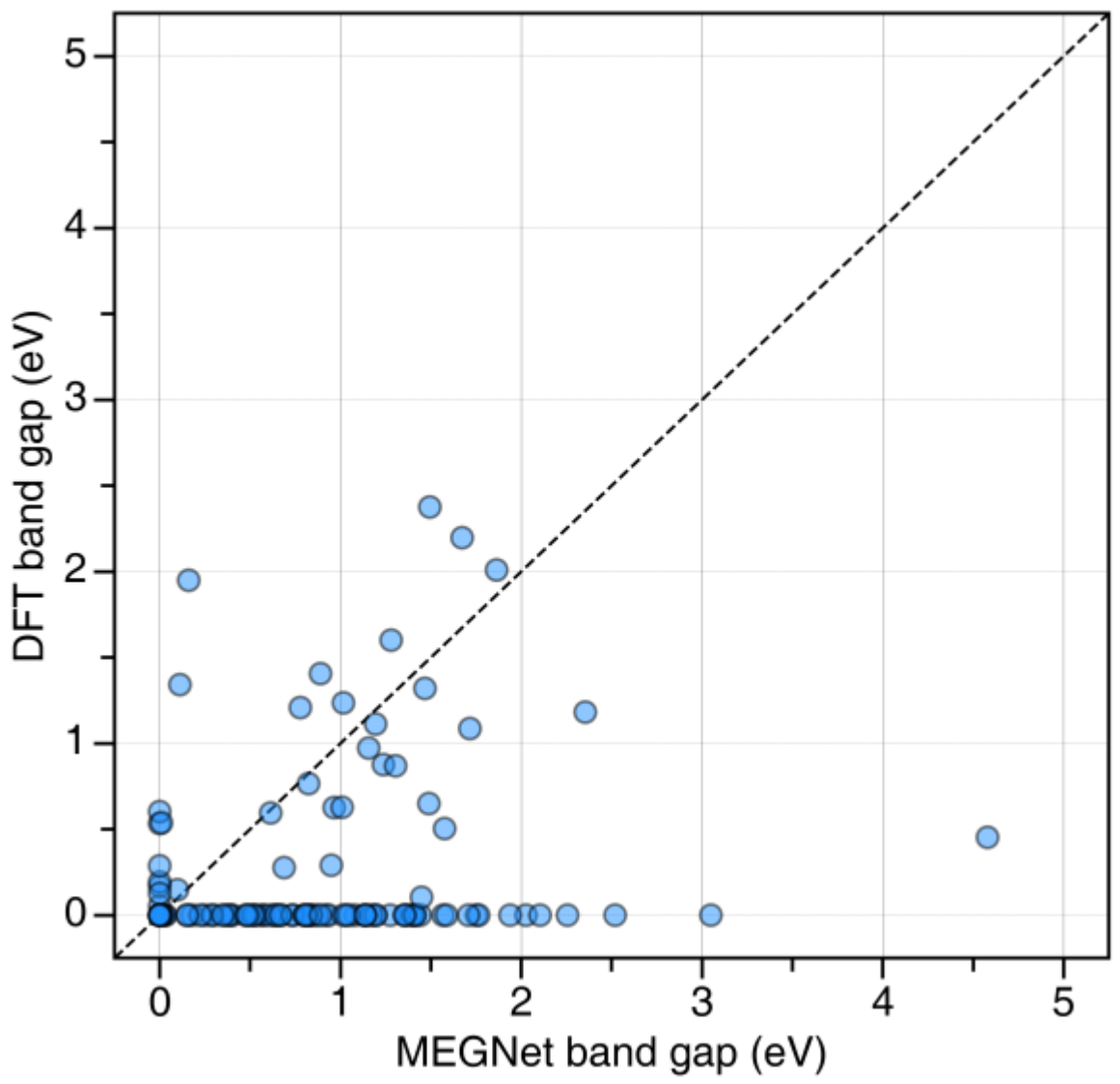

**Supplementary Figure 2 | Band gap comparison for generated crystal structures.** The DFT band gaps found in **Supplementary Table 1**. **Supplementary Table 2**

**Supplementary Table 1 | Symmetry, band gaps and stability of 35 semiconductors out of 135 inputs**

| ID | Formula | $N_{sites}$ | Space group | $E_{hull}$ (eV/atom)[a] | $E_g^{WyCryst+}$ (eV) | $E_g^{DFT}$ (eV) | Phonon[b] |
|---|---|---|---|---|---|---|---|
| 48 | $Zr_8Mn_4O_{24}$ | 36 | 205 (Pa-3) | **0.03** | 1.83 | 2.20 | Stable |
| 64 | $P_4Ir_4Se_{16}$ | 24 | 198 ($P2_13$) | 0.32 | 0.85 | 0.15 | Unstable |
| 80 | $K_{12}Bi_4Se_{20}$ | 36 | 198 ($P2_13$) | 0.64 | 1.39 | 0.77 | Unstable |
| 84 | $Zr_4Rh_{16}S_{20}$ | 40 | 198 ($P2_13$) | 0.27 | 0.64 | 0.05 | Unstable |
| 102 | $Cs_2V_2I_8$ | 12 | 194 ($P6_3$/mmc) | 0.09 | 0.62 | 0.17 | Unstable |
| 132 | $Cs_2Tl_6Te_{10}$ | 18 | 194 ($P6_3$/mmc) | 0.48 | 0.51 | 0.28 | Unstable |
| 202 | $K_8Mn_2S_6$ | 16 | 186 ($P6_3mc$) | 0.21 | 1.25 | 0.19 | Unstable |
| 203 | $Na_6Mn_2Se_8$ | 16 | 186 ($P6_3mc$) | **0.06** | 1.24 | 0.59 | Stable |
| 239 | $Rb_{12}Hg_4S_{10}$ | 26 | 186 ($P6_3mc$) | **0.10** | 1.89 | 1.41 | Stable |
| 251 | $K_{12}Hg_8S_{10}$ | 30 | 186 ($P6_3mc$) | 0.33 | 1.69 | 0.53 | Unstable |
| 352 | $Pt_3Pb_3F_{24}$ | 30 | 166 (R-3m) | 0.16 | 2.06 | 0.50 | Unstable |
| 378 | $Cs_6Pt_9Se_{21}$ | 36 | 166 (R-3m) | **0.06** | 1.01 | 0.13 | Stable |
| 385 | $Pt_3Pb_6F_{24}$ | 33 | 166 (R-3m) | 0.17 | 2.01 | 2.01 | Unstable |
| 556 | $Cd_6Sn_6Br_{36}$ | 48 | 161 (R3c) | 0.10 | 1.40 | 1.60 | Unstable |
| 566 | $Tl_9Sb_3O_9$ | 21 | 160 (R3m) | 0.17 | 1.63 | 1.24 | Unstable |
| 579 | $Fe_3Bi_9O_{18}$ | 30 | 160 (R3m) | 0.19 | 1.77 | 1.95 | Unstable |
| 584 | $Tl_3As_9S_{15}$ | 27 | 160 (R3m) | 0.10 | 1.29 | 1.34 | Unstable |
| 586 | $Tl_9As_9S_{12}$ | 30 | 160 (R3m) | 0.22 | 1.30 | 0.29 | Stable |
| 597 | $Y_9Ag_3S_{27}$ | 39 | 160 (R3m) | 0.09 | 0.87 | 1.09 | Unstable |
| 612 | $Ge_3Pb_9S_{36}$ | 48 | 160 (R3m) | 0.36 | 1.29 | 0.63 | Unstable |
| 669 | $Hg_6Pb_3F_{18}$ | 27 | 148 (R-3) | 0.18 | 1.20 | 2.38 | Unstable |
| 693 | $Cd_9P_6Se18$ | 33 | 148 (R-3) | 0.20 | 0.95 | 0.88 | Stable |
| 703 | $Cd_{12}Ge_{12}O_{18}$ | 42 | 148 (R-3) | 0.23 | 0.75 | 0.29 | Stable |
| 704 | $Co_9Pt_3F_{36}$ | 48 | 148 (R-3) | 0.20 | 0.97 | 0.60 | Unstable |
| 713 | $Al_6Ge_5S_{11}$ | 22 | 143 (P3) | 0.20 | 2.16 | 0.87 | Stable |
| 786 | $Rb_8V_8Br_{28}$ | 44 | 64 (Cmce) | 0.16 | 0.97 | 0.65 | Unstable |
| 851 | $K_6Au_{14}S_{12}$ | 32 | 55 (Pbam) | 0.14 | 0.97 | 0.97 | Unstable |
| 880 | $Na_4Ag_4S_8$ | 16 | 33 ($Pna2_1$) | 0.10 | 1.36 | 1.11 | Stable |
| 898 | $Cr_4Hg_4Cl_{16}$ | 24 | 19 ($P2_12_12_1$) | 0.08 | 0.73 | 1.21 | Stable |
| 905 | $Al_4Bi_8Br_{36}$ | 48 | 15 (C2/c) | 0.18 | 1.89 | 1.18 | Unstable |
| 925 | $Ba_1Br_3N_2$ | 6 | 10 (P2/m) | 0.21 | 1.08 | 0.45 | Unstable |
| 931 | $Cs_2P_3Se_8$ | 13 | 6 (Pm) | 0.12 | 1.38 | 0.10 | Unstable |
| 934 | $K_{12}Cd_4O_{10}$ | 26 | 5 (C2) | 0.10 | 1.30 | 1.32 | Stable |
| 942 | $K_1Tl_3O_4$ | 8 | 2 (P-1) | 0.09 | 0.47 | 0.54 | Stable |
| 944 | $Na_3In_3S_4$ | 10 | 2 (P-1) | 0.15 | 1.59 | 0.63 | Unstable |

[a] DFT calculated results; Phonon stable semiconductors with $E_{hull}$ <= 0.10 eV/atom are in bold.

[b] Estimated from density functional perturbation theory (DFPT) calculations at the Γ q-point.

**Supplementary Table 2 | Similar structures in the Materials Project to the new 8 semiconductors**

| ID | Space group | Formula | $E_g$ (eV) | MP-ID[a] | Space group | Formula[b] | ICSD ID | $E_g$ (eV)[c] |
|---|---|---|---|---|---|---|---|---|
| 378 | 166 (R-3m) | $Cs_6Pt_9Se_{21}$ | 0.13 | mp-573316 | 166 (R-3m) | $Cs_6Pt_{12}Se_{18}$ | 69440 | 1.05 |
| 703 | 148 (R-3) | $Cd_{12}Ge_{12}O_{18}$ | 0.29 | mp-8275 | 148 (R-3) | $Cd_6Ge_6O_{18}$ | 30971 | 1.41 |
| 586 | 160 (R3m) | $Tl_9As_9S_{12}$ | 0.29 | mp-9791 | 160 (R3m) | $Tl_9As_3S_9$ | 100292 | 1.22 |
| 203 | 186 (P6₃mc) | $Na_6Mn_2Se_8$ | 0.59 | mp-14780 | 186 (P6₃mc) | $Na_{12}Mn_2Se_8$ | 65449 | 1.13 |
| | | | | mp-10232 | 156 (P3m1) | $NaMnSe_2$ | 50818 | 0 (1.06)[d] |
| | | | | mp-29745 | 15 (C2/c) | $Na_{16}Mn_{16}Se_{24}$ | 50820 | 0 (1.60)[e] |
| 713 | 143 (P3) | $Al_6Ge_5S_{11}$ | 0.87 | — | — | — | — | — |
| 693 | 148 (R-3) | $Cd_9P_6Se_{18}$ | 0.88 | mp-1079559 | 148 (R-3) | $Cd_6P_6Se_{18}$ | 620234 | 1.05 |
| 239 | 186 (P6₃mc) | $Rb_{12}Hg_4S_{10}$ | 1.41 | mp-1190842 | 186 (P6₃mc) | $Rb_{12}Hg_2S_8$ | 639158 | 1.60 |
| 48 | 205 (Pa-3) | $Zr_8Mn_4O_{24}$ | 2.20 | mp-754513 | 161 (R3c) | $Zr_6Mn_6O_{18}$ | — (0.04)[f] | 2.70 |
| | | | | mp-763464 | 1 (P1) | $Zr_4MnO_9$ | — (0.05)[f] | 2.66 |

[a] The identifier (ID) in the Materials Project.
[b] The formulae are in the reduced form.
[c] The band gap is as-queried, calculated at the GGA(+$U$)-PBE level.
[d] DFT calculated from ref [1], for which the MP gives zero gaps.
[e] DFT+$U$ calculated from ref [2], where another analog with reduced formula of $Na_4Mn_6Se_8$ is presented with C2/m symmetry and band gap of 1.59 eV, and the measured optical band gaps for $Na_{16}Mn_{16}Se_{24}$ and $Na_4Mn_6Se_8$ are 2.03 and 2.04 eV, respectively.
[f] The energy above the convex hull given by the Materials Project website (Accessed date: 2024 Sept 12).

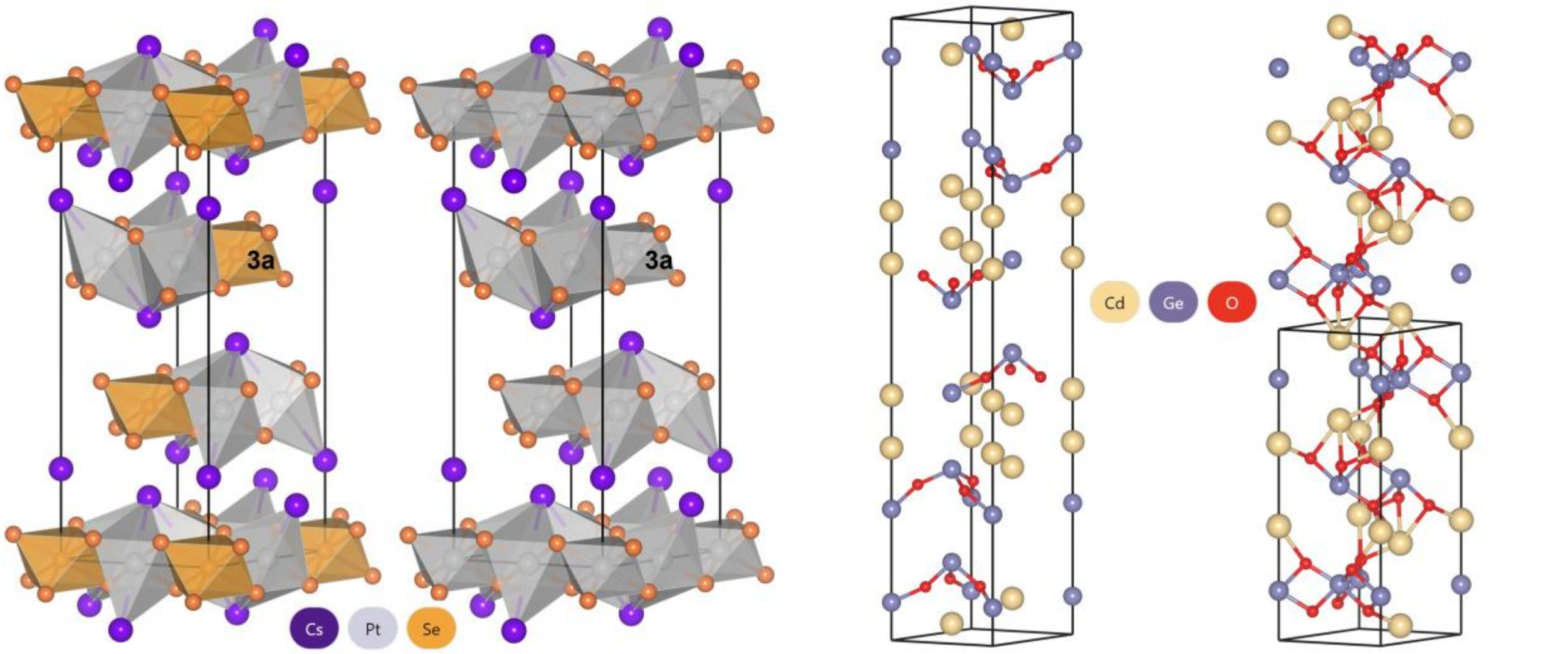

| (378) | mp-573316 |
|---|---|
| R-3m (166) | R-3m (166) |
| $Cs_6Pt_9Se_{21}$ | $Cs_6Pt_{12}Se_{18}$ |
| Cs: 6c (0, 0, 0.21)<br>Pt: 9e (1/2, 0, 0)<br>**Se1: 3a (0, 0, 0)**<br>Se2: 18h (0.49, -0.49, 0.27) | Cs: 6c (0, 0, 0.20)<br>**Pt1: 3a (0, 0, 0)**<br>Pt2: 9e (1/2, 0, 0)<br>Se: 18h (0.49, -0.49, 0.27) |

| (703) | mp-8275 |
|---|---|
| R-3 (148) | R-3 (148) |
| $Cd_{12}Ge_{12}O_{18}$ | $Cd_6Ge_6O_{18}$ |
| Cd1: 6c (0, 0, 0.31)<br>**Cd2: 6c (0, 0, 0.39)**<br>Ge1: 6c (0, 0, 0.06)<br>**Ge2: 6c (0, 0, 0.21)**<br>O: 18f (0.32, 0.27, 0.25) | Cd: 6c (0, 0, 0.37)<br>Ge: 6c (0, 0, 0.16)<br>O: 18f (0.34, 0.06, 0.10) |

**Supplementary Figure 3 | Structure comparison between WyCryst+-designed structures and those analogs with the same space group from the Materials Project.** Sequentially, the Wyckoff positions involve (1) substitution/replacement, (2) addition, (3) swap and addition, (4) deletion, (5) addition, (6) addition, respectively.

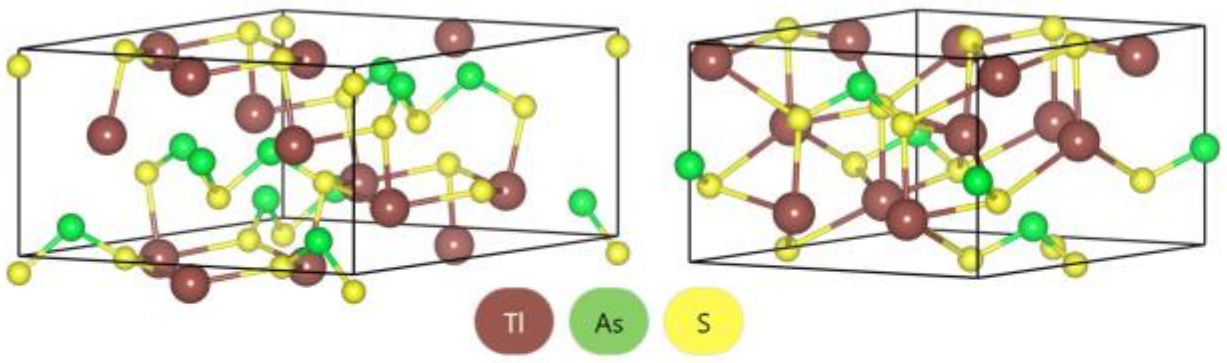

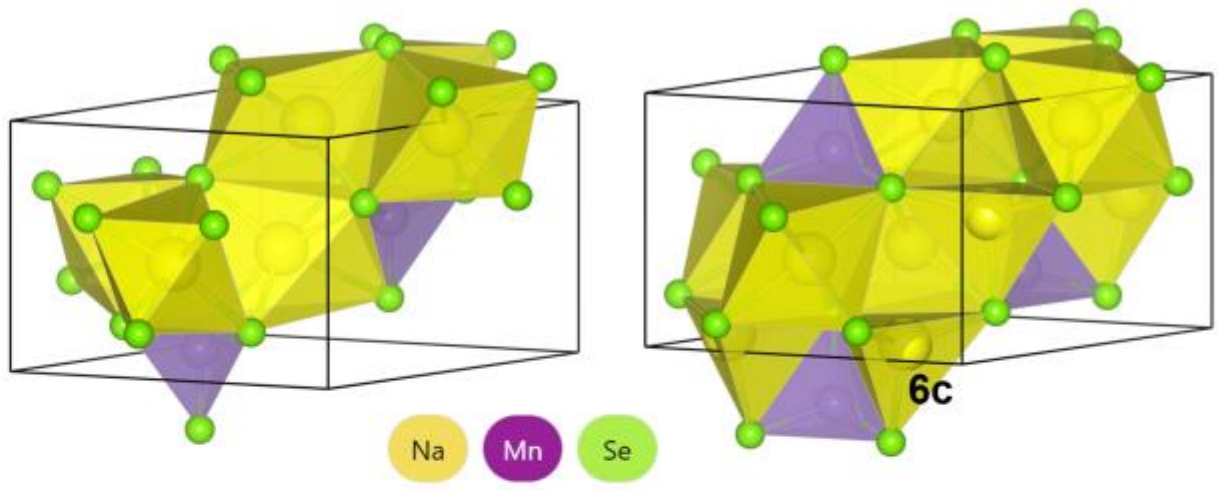

| (586) | mp-9791 |
| --- | --- |
| R3m (160) | R3m (160) |
| $Tl_9As_9S_{12}$ | $Tl_9As_3S_9$ |
| Tl: 9b (0.19, -0.19, 0.24)<br>**As: 9b (0.91, -0.91, 0.13)**<br>S1: 9b (0.82, -0.82, 0.97)<br>**S2: 3a (0, 0, 0.92)** | Tl: 9b (0.80, -0.80, 0.21)<br>As: 3a (0, 0, 0.44)<br>S: 9b (0.12, -0.12, 0.30) |

| (203) | mp-14780 |
| --- | --- |
| $P6_3mc$ (186) | $P6_3mc$ (186) |
| $Na_6Mn_2Se_8$ | $Na_{12}Mn_2Se_8$ |
| Na: 6c (0.48, -0.48, 0.95)<br>Mn: 2b (1/3, 2/3, 0.57)<br>Se1: 6c (0.20, 0.20, 0.67)<br>Se2: 2b (1/3, 2/3, 0.24) | Na1: 6c (0.47, -0.47, 0.87)<br>**Na2: 6c (0.15, -0.15, 0.54)**<br>Mn: 2b (1/3, 2/3, 0.25)<br>Se1: 6c (0.19, 0.19, 0.14)<br>Se2: 2b (1/3, 2/3, 0.60) |

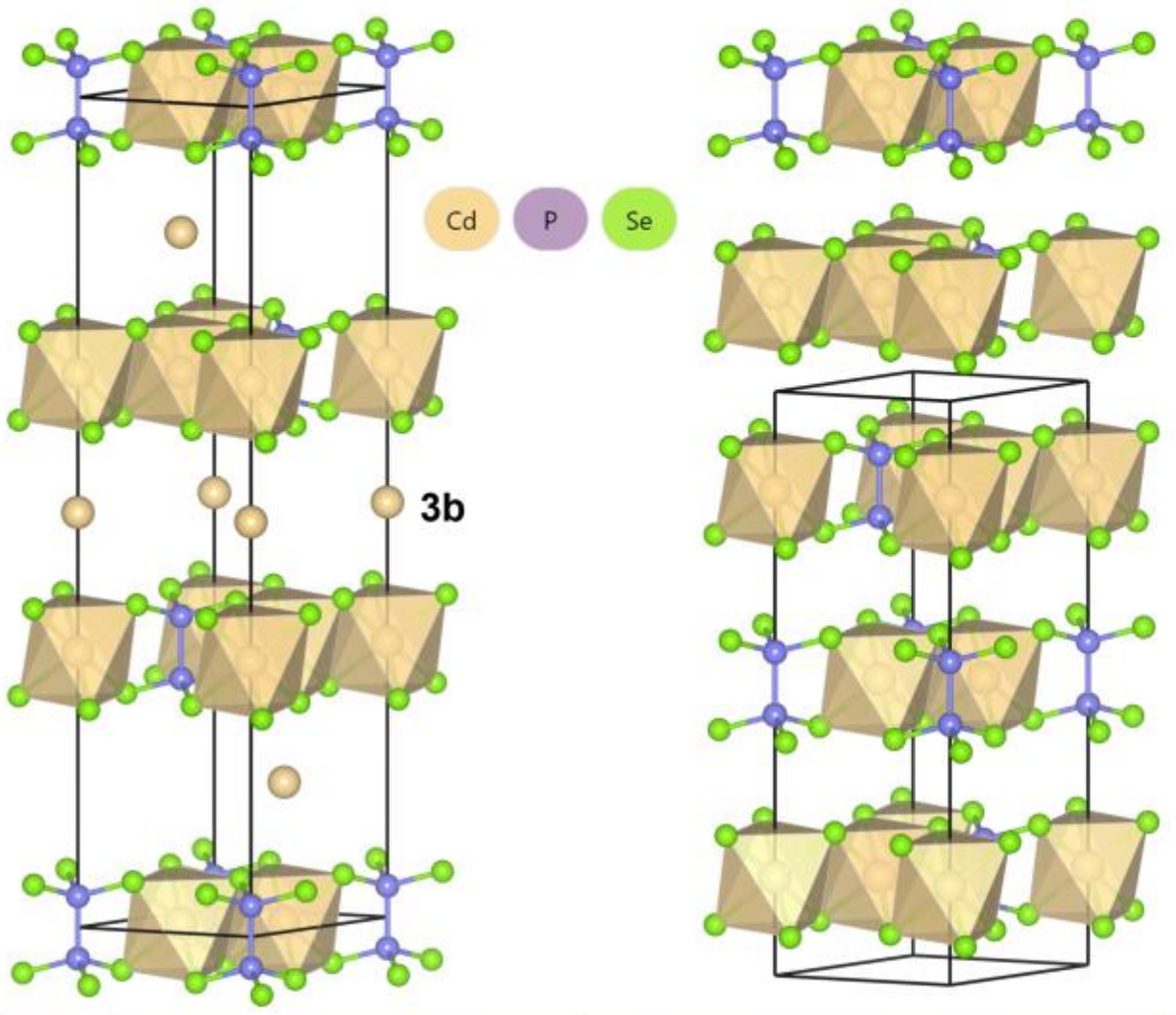

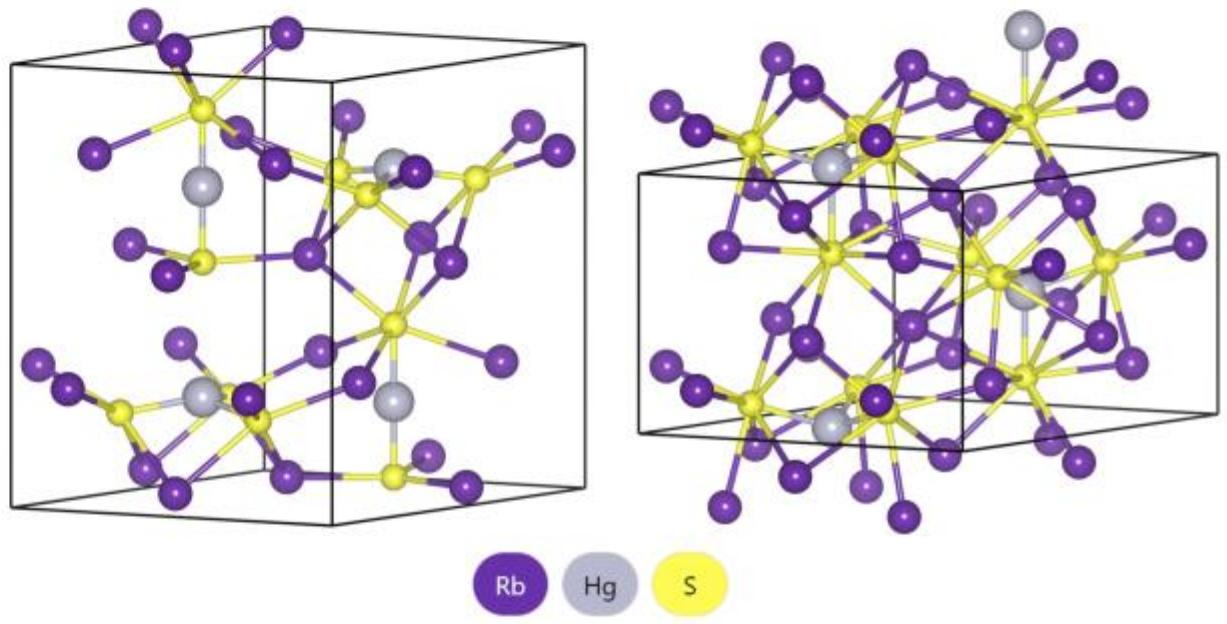

| (693) | mp-1079559 |
| --- | --- |
| R-3 (148) | R-3 (148) |
| $Cd_9P_6Se_{18}$ | $Cd_6P_6Se_{18}$ |
| Cd1: 6c (0, 0, 0.33)<br>**Cd2: 3b (0, 0, 1/2)**<br>P: 6c (0, 0, 0.04)<br>Se: 18f (0.67, 0.68, 0.06) | Cd: 6c (0, 0, 0.17)<br>P: 6c (0, 0, 0.45)<br>Se: 18f (0.34, 0.99, 0.25) |

| (239) | mp-1190842 |
| --- | --- |
| $P6_3mc$ (186) | $P6_3mc$ (186) |
| $Rb_{12}Hg_4S_{10}$ | $Rb_{12}Hg_2S_8$ |
| Rb1: 6c (0.48, -0.48, 0.55)<br>Rb2: 6c (0.15, -0.15, 0.29)<br>Hg1: 2b (1/3, 2/3, 0.21)<br>**Hg2: 3b (1/3, 2/3, 0.73)**<br>S1: 2b (1/3, 2/3, 0.04)<br>**S2: 2b (1/3, 2/3, 0.38)**<br>S3: 6c (0.19, -0.19, 0.71) | Rb1: 6c (0.47, -0.47, 0.88)<br>Rb2: 6c (0.15, -0.15, 0.21)<br>Hg: 2b (1/3, 2/3, 0.50)<br>S1: 2b (1/3, 2/3, 0.17)<br>S2: 6c (0.19, -0.19, 0.60) |

**Supplementary Figure 3** (Continued)

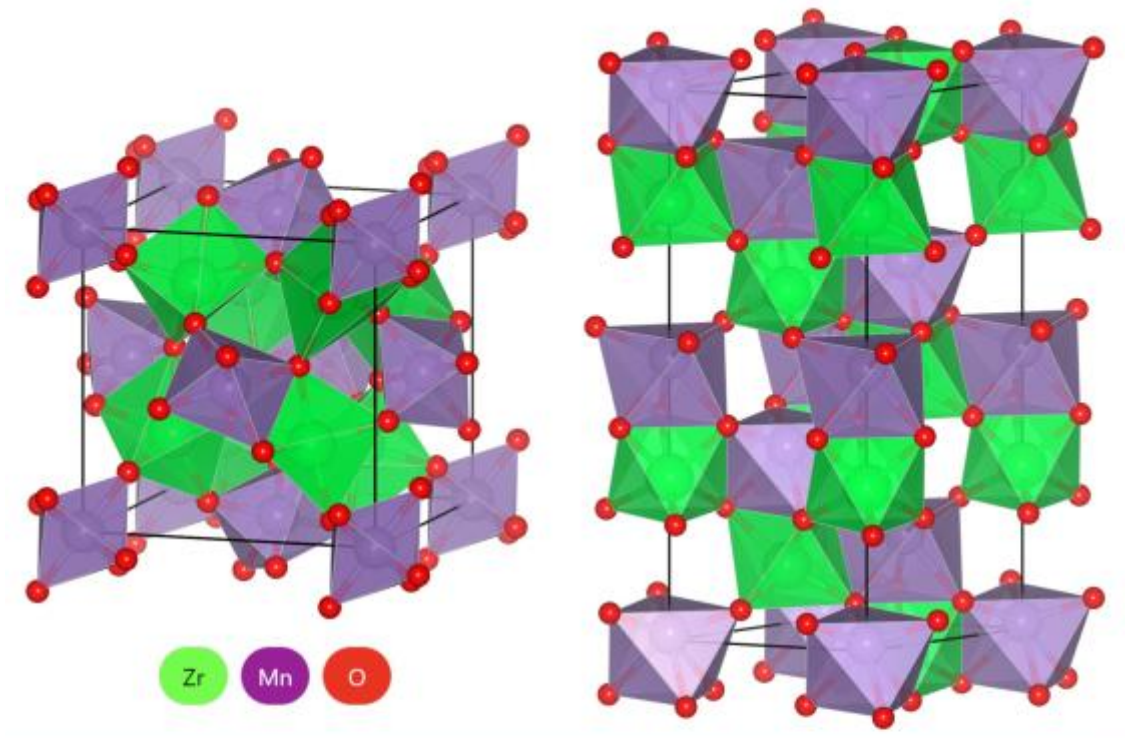

| (48) | mp-754513 |
| --- | --- |
| Pa-3 (205) | R3c (161) |
| $Zr_8Mn_4O_{24}$ | $Zr_6Mn_6O_{18}$ |
| Zr: 8c (0.21, 0.21, 0.21)<br>Mn: 4c (0, 0, 0)<br>O: 24d (0.06, 0.64, 0.31) | Zr: 6a (0, 0, 0.30)<br>Mn: 6a (0, 0, 0.01)<br>O: 18b (0.02, 0.38, 0.06) |

**Supplementary Figure 4 | Structure comparison between the WyCryst+-designed structure and the analog with the different space group from the Materials Project.** The coordination environment of Zr in the two (Zr-O polyhedra) are not the same. The former consists of distorted [$ZrO_6$] trigonal prisms while the latter distorted [$ZrO_6$] octahedra.

## Source data

Source data for Figure 4